\documentclass[aps,pra,reprint,superscriptaddress,longbibliography,nofootinbib]{revtex4-2}

\usepackage{amsmath,amssymb}
\usepackage{graphicx}
\usepackage{xcolor}
\usepackage{upgreek}
\usepackage[colorlinks=true,allcolors=blue!60!black,breaklinks=true]{hyperref}
\usepackage[capitalise]{cleveref}

\crefname{figure}{Fig.}{Figs.}
\Crefname{figure}{Figure}{Figures}
\crefname{equation}{Eq.}{Eqs.}
\Crefname{equation}{Equation}{Equations}
\crefname{section}{Section}{Sections}
\crefname{subsection}{Section}{Sections}
\crefname{appendix}{Appendix}{Appendices}
\crefname{table}{Table}{Tables}

\graphicspath{{figures/}}

\newlength{\figwidth}
\begin{document}

\title{Enhancing optomechanical oscillation amplitude and frequency comb bandwidth by intensity-modulated driving}

\author{Menno H. Jansen}
\affiliation{Center for Nanophotonics, AMOLF, 1098XG Amsterdam, The Netherlands}
\author{Ewold Verhagen}
\affiliation{Center for Nanophotonics, AMOLF, 1098XG Amsterdam, The Netherlands}
\affiliation{Department of Applied Physics and Science Education, Eindhoven University of Technology, 5612AP Eindhoven, The Netherlands}
\email{verhagen@amolf.nl}

\begin{abstract}
  Self-sustained oscillations in optomechanical systems are attractive as compact, low-noise sources of GHz signals and optical frequency combs for integrated photonic circuits. For these applications, the achievable oscillation amplitude directly sets clock power and comb bandwidth. This work studies the maximum amplitude that can be reached through combinations of cavity backaction and modulated optical driving, with realistic resources. For an optically driven mechanical resonator, the oscillation amplitude saturates at a level set by the finite cavity bandwidth. We show that modulating the input intensity near the mechanical resonance frequency raises the achievable amplitude toward this limit at equal mean input power and cavity linewidth. We use a cycle-averaged energy-balance analysis to identify the added power that shifts the stable periodic orbit to larger amplitudes. Numerical integration of the full equations of motion confirms the enhancement across two decades of cooperativity and sideband-resolution ratios from 0.1 to 10. The relative increase in amplitude is largest near the self-oscillation threshold and in the sideband-unresolved regime, enabling threshold-free large-amplitude oscillation where backaction alone fails to do so. The enhanced amplitude leads to a larger modulation depth of the optical field, enabling broader frequency combs, with improved stability of the comb spacing. Intensity-modulated driving thus offers power-constrained platforms a simple open-loop route to larger mechanical amplitudes and enhanced comb bandwidth.
\end{abstract}

\maketitle

\section{Introduction}
Radiation pressure in an optomechanical cavity couples optical and mechanical degrees of freedom, enabling optical damping or amplification of mechanical motion \cite{aspelmeyer2014review}. While cooling toward the quantum ground state has dominated attention \cite{jasperchanLaserCoolingNanomechanical2011,teufelSidebandCoolingMicromechanical2011}, the amplification branch drives the resonator into large-amplitude self-sustained oscillation. This underpins a distinct set of applications, such as the study of chaos \cite{carmonChaoticQuiveringMicronScaled2007,navarro-urriosNonlinearDynamicsChaos2017} and microwave photonics \cite{capmanyMicrowavePhotonicsCombines2007}. It also provides a source of coherent phonons on a chip for quantum and classical information technology \cite{grudininPhononLaserAction2010, safavi-naeiniControllingPhononsPhotons2019}.

Self-sustained oscillation is particularly attractive for signal generation: the oscillating resonator periodically modulates the cavity it is coupled to, imprinting its GHz-range motion on the transmitted light. This makes optomechanical oscillators promising on-chip clocks for telecommunication and positioning applications, offering an attractive combination of frequency stability, phase noise, and compactness \cite{luanIntegratedLowPhase2014,mercadeMicrowaveOscillatorFrequency2018,ghorbelOptomechanicalGigahertzOscillator2019}. Beyond a single tone, sufficiently strong oscillation can generate an optical frequency comb with a spacing of the mechanical frequency \cite{miriOptomechanicalFrequencyCombs2018}, which is useful for a host of applications, including spectroscopy, wavelength (de)multiplexing, and optical ranging \cite{fortier20YearsDevelopments2019,diddamsOpticalFrequencyCombs2020}. For both, the oscillation amplitude is a central figure of merit, as it sets the modulation depth of the output field and thereby the power of the clock signal and the span of the comb \cite{huGenerationOpticalFrequency2021}.

\begin{figure*}[t]
  \centering
  \includegraphics[
    width=\figwidth
  ]{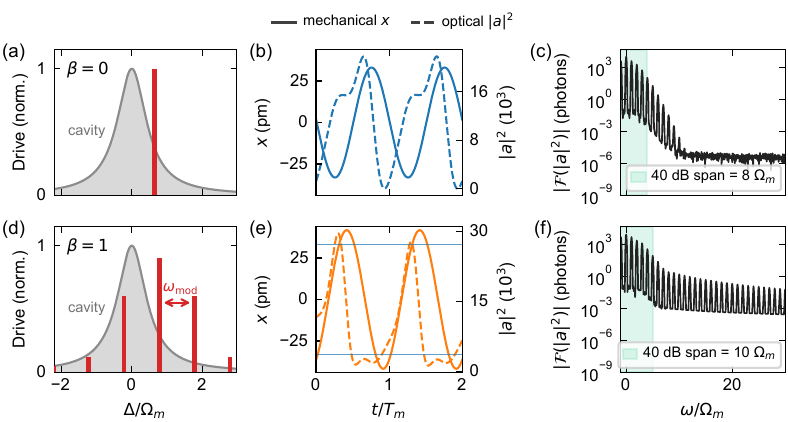}
  \caption{
    \textbf{Backaction-only versus intensity-modulated driving.} Top row, backaction-only driving ($\beta=0$); bottom row, full-depth intensity modulation ($\beta=1$), both at the same input power and cavity linewidth ($C_0\approx15$, $\kappa/\Omega_m=1$) and each shown at its optimal detuning. (a),(d) Optical drive spectrum in the cavity frame: the shaded Lorentzian is the cavity response, red lines the drive, placed at the carrier detuning $\Delta/\Omega_m$ and, under modulation, flanked by sidebands spaced by $\omega_\mathrm{mod}$. The sideband heights in (d) follow from the square-root amplitude $\sqrt{1+\beta\cos\omega_\mathrm{mod}t}$. (b),(e) Resulting steady-state traces over two mechanical periods: mechanical displacement $x$ (solid, left axis) and intracavity photon number $|a|^2$ (dashed, right axis). Blue horizontal lines in (e) mark the mechanical amplitude of (b) for comparison; modulation increases the oscillation amplitude at equal mean power. (c),(f) Spectrum of $|a|^2(t)$, versus harmonic order $\omega/\Omega_m$. The shaded band shows the $40\,\mathrm{dB}$ contiguous span.
    \label{fig:intro_schematic}
  }
\end{figure*}

The conventional way to manipulate mechanical dissipation in an optomechanical system is dynamical backaction. Here the optical force on the mechanics depends on the mechanical position, with a delay set by the photon lifetime in the cavity. In this mechanism, the detuning of the laser driving the system relative to the optical cavity determines whether it is damping or amplifying the mechanical motion. By driving blue-detuned, as schematically displayed in \cref{fig:intro_schematic}(a), the mechanical motion can be amplified into a self-oscillating state. \Cref{fig:intro_schematic}(b) shows an example of the oscillation of mechanical displacement and optical intensity in this state, calculated with the methods described below. The spectrum of the optical field then features a comb of frequencies, as shown in \cref{fig:intro_schematic}(c). The amplitude of this state saturates when the cycle-averaged gain is equal to the loss. This limit is essentially set by the finite bandwidth of the cavity \cite{pootBackactionLimitsSelfsustained2012}.

Where the resonator saturates depends on the detuning, since the zero-gain contour of the attractor diagram \cite{marquardtDynamicalMultistabilityInduced2006} runs to larger amplitude at larger blue detuning. Sweeping the detuning after the onset of oscillation makes the resonator follow this contour to an amplitude 500 times the thermal value \cite{butersExperimentalExplorationOptomechanical2015}. This state is not reachable from the thermal state at the final detuning and must be prepared along a path. Actively modulating the laser frequency to track the optimal detuning makes this steering continuous \cite{parsaFeedbackEnhancedPhonon2024}. A modulated pump can likewise steer the resonator onto an otherwise inaccessible attractor, sustaining self-oscillation even at red detuning \cite{krauseNonlinearRadiationPressure2015}. Furthermore, two-tone driving has been shown to enhance the amplitude of a self-oscillating mechanical resonator, although it was thus far studied in a limited parameter range \cite{guOpticalFrequencyComb2024, wuAmplitudePhaseLocking2022}.

An alternate approach is modulating the input intensity near the mechanical resonance frequency, as represented in \cref{fig:intro_schematic}(d). It is the symmetric extension of two-tone driving, and needs only a modulator on the pump rather than a second laser phase-locked to the first. The drive carries phase-coherent sidebands about the carrier, whose beat with the mean field injects power into the mechanical mode at the drive frequency \cite{weisOptomechanicallyInducedTransparency2010, safavi-naeiniElectromagneticallyInducedTransparency2011, agarwalElectromagneticallyInducedTransparency2010}. At equal mean power, this can raise the oscillation amplitude beyond the backaction limit, as shown by the orange line in \cref{fig:intro_schematic}(e) exceeding the blue bounds of the unmodulated amplitude. We derive the conditions for this enhancement in \cref{subsec:energyBalance} and map its extent numerically in \cref{subsec:AmpEnhancement}. Because the drive tone then sets the oscillation frequency, this scheme also injection-locks the oscillator. This provides frequency tuning, removes the influence of the optomechanical spring shift on the comb spacing, and reduces phase noise \cite{hossein-zadehObservationInjectionLocking2008, zhengFeedbackHarmonicLocking2013}.

Pump intensity modulation has indeed been used to injection-lock and synchronize optomechanical oscillators across several platforms \cite{shlomiSynchronizationOptomechanicalCavity2015, arreguiInjectionLockingOptomechanical2021, rodriguesOptomechanicalSynchronizationMultioctave2021, navarro-urriosGiantInjectionlockingBandwidth2022}. These works establish that modulation can drive, lock, and tune an oscillator, but each characterizes a single device at a single operating point. Whether modulation, or a combination of modulation and cavity backaction, consistently outperforms backaction-induced self-oscillation at the same mean power has not been assessed. That requires evaluating both strategies at their respective optimal detunings across the parameter space. We make this comparison for a resonator started from its thermal state at fixed detuning, the operating condition of a device that is switched on rather than steered. Such detuning sweeps remain available on top of the modulation. Moreover, the energy balance derived below shows that, to first order in the modulation depth $\beta$ and with the drive injection-locked, the modulation adds power at every amplitude and detuning. The outward shift it produces therefore applies to every branch of the attractor diagram, not only the one reached from the thermal state, as we show in Figs.~\ref{fig:energy}(c) and \ref{fig:energy}(d).

In this work, we carry out such a systematic comparison, showing that for a range of realistic input powers and cavity linewidths, intensity-modulated driving achieves a larger amplitude than relying on dynamical backaction. This larger amplitude also translates into wider frequency combs, an example of which is shown in Figs.~\ref{fig:intro_schematic}(c) and \ref{fig:intro_schematic}(f).

\section{Model and equations of motion}
\label{sec:modelEOM}
We treat a single mechanical mode of effective mass $m$, resonance frequency $\Omega_m$ and damping rate $\Gamma_m$, dispersively coupled to a single optical cavity mode with total decay rate $\kappa$ and external coupling rate $\kappa_\mathrm{ex}$. The cavity field amplitude $a$ is normalized so that $|a|^2$ is the intracavity photon number. In the frame rotating at the laser frequency $\omega_L$ the equations of motion are
\begin{align}
  m \ddot{x} + m \Gamma_m \dot{x} + m \Omega_m^2 x &= F_\mathrm{rp} + F_\mathrm{th},
  \label{eq:mech}\\
  \dot a &= \bigl(i(\Delta + G x) - \tfrac{\kappa}{2}\bigr) a + \sqrt{\kappa_\mathrm{ex}}\,a_\mathrm{in},
  \label{eq:cav}
\end{align}
where $G = -\partial\omega_c/\partial x$ is the optical frequency change per unit displacement, $\Delta = \omega_L-\omega_c$ is the bare detuning, and $F_\mathrm{rp}=\hbar G|a|^2$ is the radiation-pressure force. The mechanical damping is accompanied by a fluctuating thermal force $F_\mathrm{th}$ from the same bath, with (symmetrized) correlator $\langle F_\mathrm{th}(t)F_\mathrm{th}(t')\rangle = 2m\Gamma_m\hbar\Omega_m\bigl(n_\mathrm{th}+\tfrac12\bigr)\,\delta(t-t')$, where $n_\mathrm{th}=\bigl[\exp(\hbar\Omega_m/k_BT)-1\bigr]^{-1}$ is the thermal occupation of the bath at temperature $T$. The input photon flux is $|a_\mathrm{in}|^2 = P_{\mathrm{in}}/\hbar\omega_L$. The drive is intensity modulated with modulation depth $\beta$, $P_{\mathrm{in}}(t)=P_0(1+\beta \cos \omega_\mathrm{mod} t)$, so $a_\mathrm{in}=\bar{a}_\mathrm{in} \sqrt{1 + \beta \cos \omega_\mathrm{mod}t}$ with $\bar{a}_\mathrm{in}=\sqrt{P_0/\hbar\omega_L}$. When the mechanical resonator is driven to large amplitude, either by dynamical backaction or direct driving, we assume that it settles on a sinusoidal stable periodic orbit: $x(t)=\bar x+A\cos\Omega t$ \cite{marquardtDynamicalMultistabilityInduced2006, ludwigOptomechanicalInstabilityQuantum2008}. Through the $iGx\,a$ term in \cref{eq:cav}, this motion phase-modulates the intracavity field with depth
\begin{equation}
  \mu = \frac{GA}{\Omega},
  \label{eq:mu}
\end{equation}
the optical peak phase excursion per mechanical cycle \cite{marquardtDynamicalMultistabilityInduced2006, krauseNonlinearRadiationPressure2015}. Solving \cref{eq:cav} on this trajectory for the unmodulated drive $a_\mathrm{in}=\bar a_\mathrm{in}$, which we do in \cref{sup:FirstOrderPower}, shows how $\mu$ shapes the intracavity field,
\begin{equation}
  a(t)=\sqrt{\kappa_\mathrm{ex}}\,\bar a_\mathrm{in}\,e^{i\mu\sin\Omega t}\sum_{n=-\infty}^{\infty}\frac{J_n(\mu)\,e^{-in\Omega t}}{\tfrac\kappa2-i(\bar\Delta+n\Omega)},
  \label{eq:a0}
\end{equation}
where $\bar\Delta=\Delta+G\bar x$ is the detuning renormalized by the static displacement. The Bessel weights $J_n(\mu)$, which become small for $|n|\gtrsim\mu$, spread the field over sidebands at multiples of $\Omega$, each filtered by the cavity response at $\bar\Delta+n\Omega$. Through these weights, $\mu$ governs both the saturation of the backaction power and the width of the generated frequency comb, as we will see in \cref{subsec:combscaling}.

Multiplying \cref{eq:mech} by $\dot x$ and integrating over one mechanical
period $T$, the kinetic and potential energy terms average to zero, and we obtain the balance between energy dissipated by damping and energy injected by radiation pressure \cite{marquardtDynamicalMultistabilityInduced2006, ludwigOptomechanicalInstabilityQuantum2008, metzgerSelfInducedOscillationsOptomechanical2008, butersExperimentalExplorationOptomechanical2015, krauseNonlinearRadiationPressure2015}:

\begin{align}
  \begin{split}
    W_{\mathrm{diss}}&=W_{\mathrm{rp}}, \\
    m \Gamma_m \int \dot{x}^2 dt &= \int F_{\mathrm{rp}} \dot{x} dt.
  \end{split}
\end{align}

Inserting the sinusoidal assumption for the mechanical amplitude, and ignoring higher harmonics due to the high quality factor of the resonator, the cycle-averaged dissipated and injected powers become

\begin{align}
  \begin{split}
    P_{\mathrm{diss}}(A) &= \frac{W_{\mathrm{diss}}}{T} = \tfrac12 m\Gamma_m\Omega^2 A^2, \\
    P_{\mathrm{rp}}(A,\beta) &= \frac{W_{\mathrm{rp}}}{T} = \hbar G\,\bigl\langle |a|^2\dot x\bigr\rangle,
  \end{split}
  \label{eq:powers}
\end{align}
where $\langle\cdot\rangle$ denotes the average over one mechanical period. The dissipated power follows directly from $\dot x=-A\Omega\sin\Omega t$, whereas the injected power depends on $A$ through the phase-modulation depth $\mu$ of the intracavity field, as we can see in \cref{eq:a0}, and on $\beta$ through the drive.

The thermal force also does work on the resonator, injecting on average $P_\mathrm{th}=\langle F_\mathrm{th}\dot x\rangle=\Gamma_m\hbar\Omega_m\bigl(n_\mathrm{th}+\tfrac12\bigr)$, the power that sustains the equilibrium fluctuations $x_\mathrm{th}^2=(2n_\mathrm{th}+1)x_\mathrm{zpf}^2$ in the absence of optical driving. Relative to the dissipation in the periodic steady state this contribution is $P_\mathrm{th}/P_\mathrm{diss}=2x_\mathrm{th}^2/A^2$, which is below $10^{-4}$ for the tens-of-picometer amplitudes considered here. We therefore drop $F_\mathrm{th}$ from the cycle-averaged energy balance. The bath instead matters in two other ways. It seeds the growth of self-oscillation out of the unstable zero-amplitude state. It also drives residual phase and amplitude diffusion about the stable periodic orbit, against which the injection-locked phase $\psi^\star$ discussed below must be maintained. Combining the injected and dissipated powers, we define the net-gain function

\begin{equation}
  \mathcal{G}(A,\beta) = P_{\mathrm{rp}}(A,\beta)-P_{\mathrm{diss}}(A).
  \label{eq:gain}
\end{equation}

We can find the amplitudes $A_*$ of stable periodic orbits by evaluating its solutions such that $\mathcal{G}(A_*,\beta)=0$ and $\partial_A\mathcal{G}(A_*,\beta)<0$ \cite{marquardtDynamicalMultistabilityInduced2006}.

\subsection{Effect of modulation on energy balance}
\label{subsec:energyBalance}

To find the effect of the modulation on the amplitude of the stable periodic orbit, we start by expanding the drive in the modulation depth $\beta$. Writing the square root in $a_\mathrm{in}=\bar a_\mathrm{in}\sqrt{1+\beta\cos\omega_\mathrm{mod}t}$ as a power series gives

\begin{align}
  \begin{split}
    a_\mathrm{in} &=\bar{a}_\mathrm{in}\Bigl(1+\frac{\beta}{2}\cos\omega_\mathrm{mod}t - \frac{\beta^{2}}{8}\cos^{2}\omega_\mathrm{mod}t+\mathcal{O}(\beta^{3})\Bigr) \\
    &= \bar{a}_\mathrm{in} + \delta a_\mathrm{in}^{(1)}+\delta a_\mathrm{in}^{(2)}+\dots
  \end{split}
  \label{eq:ain-exp}
\end{align}

The series converges for $|\beta|<1$, and expresses the modulated drive as a set of sidebands about the carrier \cite{mariGentlyModulatingOptomechanical2009}. At full depth, $\beta=1$, the drive amplitude $\sqrt{1+\beta\cos\omega_\mathrm{mod}t}$ reaches the branch point of the square root once per cycle, where it vanishes. The expansion in \cref{eq:ain-exp} is used only to identify the sign-protected first-order mechanism. The full-depth mechanical amplitudes are obtained numerically from the exact drive in \cref{sec:numResults}, without expansion.

As \cref{eq:cav} is linear in $a$ for a fixed trajectory $x(t)$, the cavity field inherits the same behavior, $a=a^{(0)}+\delta a^{(1)}+\delta a^{(2)}+\dots$, where $\delta a^{(n)}$ is the cavity response to $\delta a_\mathrm{in}^{(n)}$ and scales as $\beta^{n}$. The mean-field response $a^{(0)}$ is driven by $\bar a_\mathrm{in}$ alone and, evaluated on the prescribed periodic steady state, carries the unmodulated $\beta=0$ backaction, $P_\mathrm{rp}^{(0)}(A)=P_\mathrm{rp}(A,0)$ \cite{marquardtDynamicalMultistabilityInduced2006, russoAnalyticalExplorationOptomechanical2025}.

Inserting this expansion into the injected power of \cref{eq:powers}, and using $|a|^{2}=|a^{(0)}|^{2}+2\mathrm{Re}[a^{(0)}{}^{*}\delta a]+|\delta a|^{2}$, sorts the correction by powers of $\beta$:

\begin{equation}
  \begin{split}
  P_\mathrm{rp}(A,\beta)={}&P_\mathrm{rp}^{(0)}(A)
  +\underbrace{2 \hbar G \bigl\langle \mathrm{Re} \left[a^{(0)}{}^{*} \delta a^{(1)} \right] \dot x \bigr\rangle}_{\displaystyle \Delta P_1=\mathcal{O}(\beta)}\\
  &+\underbrace{\hbar G \bigl\langle \bigl( 2 \mathrm{Re}\left[ a^{(0)}{}^{*}\delta a^{(2)} \right] + |\delta a^{(1)}|^{2} \bigr) \dot x \bigr\rangle}_{\displaystyle \Delta P_2=\mathcal{O}(\beta^{2})}\\
  &+\mathcal{O}(\beta^{3}).
  \end{split}
  \label{eq:Prp-orders}
\end{equation}

The first-order correction term $\Delta P_1$ is the beat of the mean field with the first-order sidebands. The second-order term $\Delta P_2$ collects the first-order sideband intensity $|\delta a^{(1)}|^{2}$ together with the beat of the mean field with the second-order sidebands \cite{xiongHigherorderSidebandsOptomechanically2012}. The dissipated power $P_\mathrm{diss}=\tfrac12 m\Gamma_m\Omega^2A^2$ depends on the modulation only through the oscillation frequency $\Omega$. With the drive inside the locking range derived in \cref{supp:locking}, the oscillator follows the drive, $\Omega=\omega_\mathrm{mod}\simeq\Omega_m$, while without modulation $\Omega$ differs from $\Omega_m$ only by the small optical spring shift. $P_\mathrm{diss}$ is then independent of $\beta$, and the modulation enters the net gain of \cref{eq:gain} only through $P_\mathrm{rp}$. Substituting the unmodulated gain $\mathcal{G}(A,0) = P_\mathrm{rp}^{(0)} - P_\mathrm{diss}$ into \cref{eq:gain} then yields

\begin{equation}
  \mathcal{G}(A,\beta) = \mathcal{G}(A,0) + \Delta P (A,\beta).
\end{equation}

\begin{figure*}[t]
  \centering
  \includegraphics[
    width=\figwidth
  ]{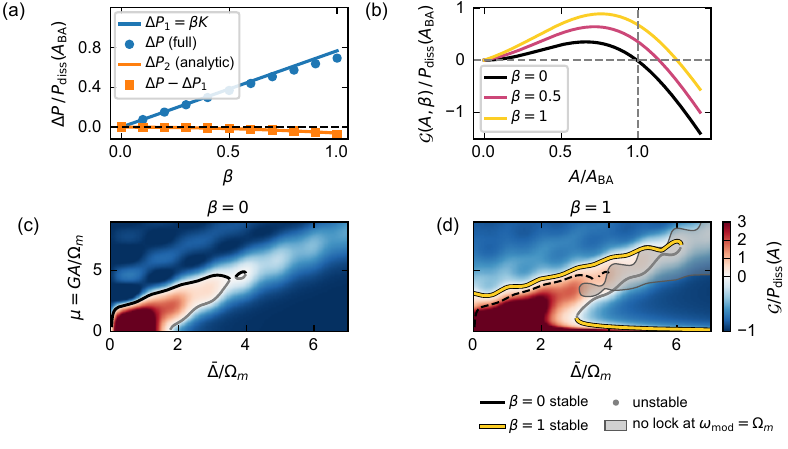}
  \caption{
    \textbf{Intensity modulation adds power to the energy balance, shifting the stable periodic orbits outward at every detuning.}\\
    (a) Power added at the backaction amplitude $A_\mathrm{BA}$ versus modulation depth $\beta$, at the optimal modulation phase. The first-order term $\Delta P_1=\beta K$ matches the full $\Delta P$ at small $\beta$, and the analytic second-order term $\Delta P_2\propto\beta^2$ accounts for the remainder $\Delta P-\Delta P_1$ up to $\beta=1$.
    (b) Net gain $\mathcal{G}(A,\beta)$ versus oscillation amplitude. Modulation moves the downward zero crossing, and with it the stable orbit, beyond $A_\mathrm{BA}$ (vertical dashed line). Panels (a) and (b) use the exact steady state of \cref{eq:gain} on $x(t)=A\cos\Omega t$, at $C_0\approx15$, $\kappa/\Omega_m=1$ and $\bar\Delta=\Omega_m$.
    (c) Attractor diagram: $\mathcal{G}$ normalized to the local $P_\mathrm{diss}(A)$, computed as in (b) for the same cavity at $P_0=500\,\upmu\mathrm{W}$ ($C_0\approx49$).
    (d) The same at $\beta=1$, with the modulation frequency tracking the free-running frequency so that the drive stays locked at the phase of maximal injected power. Dashed: the $\beta=0$ stable orbits of (c). Shaded: no phase-stable lock for a drive fixed at $\omega_\mathrm{mod}=\Omega_m$.
    \label{fig:energy}
  }
\end{figure*}

We see that the effect of the intensity modulation is to add a power $\Delta P = \Delta P_1 + \Delta P_2$ to the net gain evaluated at the backaction amplitude $A_\mathrm{BA}$. \Cref{fig:energy}(a) shows that $\Delta P_1$ dominates at small $\beta$, and that $\Delta P_2$ accounts for the remainder up to $\beta=1$.

The sign of $\Delta P_1$ is controlled by the modulation phase. In the periodic steady state, the force $\delta F=2\hbar G\,\mathrm{Re}[a^{(0)}{}^{\!*}\delta a^{(1)}]$ is periodic at $\Omega$, and only its velocity quadrature does work. With $\dot x=-A\Omega\sin\Omega t$,

\begin{equation}
  \Delta P_1=\langle\delta F\,\dot x\rangle=-A\Omega\,\mathrm{Im}\hat F_1
  =\beta\,K(A)\cos\psi,
  \label{eq:dP1-paper}
\end{equation}
where $\hat F_1$ is the fundamental Fourier component of $\delta F$ (using $\langle f\sin\Omega t\rangle=\mathrm{Im}\hat f_1$), $\psi$ is the phase of the modulation relative to the mechanical velocity, and

\begin{equation}
  K(A)=A\Omega\sqrt{\bigl[\mathrm{Im}(\Sigma_{+}+\Sigma_{-})\bigr]^{2}
  +\bigl[\mathrm{Re}(\Sigma_{+}-\Sigma_{-})\bigr]^{2}}\ \ge\ 0 ,
  \label{eq:K-paper}
\end{equation}
with the sideband sums $\Sigma_{\pm}(A,\bar\Delta,\kappa,\Omega)$ given in \cref{sup:FirstOrderPower}. This sign control is structural rather than empirical. $\Delta P_1$ is the beat of the mean field with the drive sideband and hence linear in it. As $\psi$ varies, it therefore traces a sinusoid of amplitude $\beta K$ with no constant offset, for every $A$ and detuning. The entire sign of $\Delta P_1$ therefore sits in $\cos\psi$: half of all modulation phases inject power, and the optimal phase $\psi=0$ injects the maximum $\beta K$.

Because the modulation enters only through $\Delta P$, we test its effect at the unmodulated amplitude $A_\mathrm{BA}$: if the resonator still gains energy there, the stable orbit must lie further out, as the shifted zero crossing in \cref{fig:energy}(b) shows. The unmodulated gain vanishes at $A_\mathrm{BA}$ by definition, $\mathcal{G}(A_\mathrm{BA},0)=0$, so the net gain equals the added power,
\begin{equation}
  \begin{split}
  \mathcal{G}(A_\mathrm{BA},\beta)&=\Delta P(A_\mathrm{BA},\beta)\\
  &=\beta\,K(A_\mathrm{BA})\cos\psi^\star+\Delta P_2(A_\mathrm{BA},\beta).
  \end{split}
  \label{eq:G-at-ABA}
\end{equation}
Here $\psi^\star$ is the phase between the drive and the mechanical velocity, fixed by injection locking. Within the locking range the locked state settles on the energy-injecting branch, $\cos\psi^\star>0$, so the first term is non-negative \cite{adlerStudyLockingPhenomena1946, kurokawaInjectionLockingMicrowave1973}. The exceptions lie near the lock edge, in the corner where locking already fails at some detunings. This requires the drive to capture the oscillator. Since we modulate at $\omega_\mathrm{mod}=\Omega_m$, the drive must bridge the mismatch $\delta\Omega$ between $\Omega_m$ and the free-running frequency, which the optical spring pulls away from $\Omega_m$. The locking range is the largest mismatch the drive can bridge. Within it, the oscillator adopts the drive frequency and settles at a fixed phase $\psi^\star$. Outside it, the phase slips continuously and $\Delta P_1$ averages out. In the attractor diagram of \cref{fig:energy}(d), this happens on the high-amplitude, far-blue-detuned part of the branch. The same beat that injects $\Delta P_1$ also sets this range, through its displacement quadrature \cite{amitaiSynchronizationOptomechanicalSystem2017}. As we show in \cref{supp:locking}, this gives a half-width $\Delta\Omega_\mathrm{lock}=(\Gamma_m/2)\,\beta K'/P_\mathrm{diss}$: the mechanical linewidth scaled by a power ratio that reduces to $\beta K/P_\mathrm{diss}$ for small $\mu$. Comparing $\Delta\Omega_\mathrm{lock}$ with $|\delta\Omega|$ at every sweep point, we find that the oscillator locks wherever the modulation substantially enhances the amplitude. This can be seen in \cref{fig:locking}, where the ratio $|\delta\Omega|/\Delta\Omega_\mathrm{lock}$ exceeds 1, and locking thus fails, only in the corner of high cooperativity and small $\kappa/\Omega_m$. This is also where backaction alone already approaches saturation and the enhancement from modulation is smallest.

Positive gain at $A_\mathrm{BA}$ does not yet locate the new orbit. For that, $\mathcal{G}$ must cross zero again at larger amplitude. We know that it will, because at large $A$ the two powers in \cref{eq:gain} scale oppositely with $A$: the dissipated power grows without bound, $P_\mathrm{diss}=\tfrac12 m\Gamma_m\Omega^2A^2$, while the injected power is bounded \cite{russoAnalyticalExplorationOptomechanical2025}. At large amplitude the motion phase-modulates the field deeply, $\mu\gg1$, spreading it over many sidebands whose contributions to $P_\mathrm{rp}$ increasingly cancel \cite{marquardtDynamicalMultistabilityInduced2006}. $\mathcal{G}(A,\beta)$ therefore turns negative at large $A$, and its first downward zero crossing beyond $A_\mathrm{BA}$ defines a stable periodic orbit at $A_*(\beta)>A_\mathrm{BA}$, at equal mean input power.

The argument does not rely on the particular zero crossing at $A_\mathrm{BA}$. Where the drive is locked, $\Delta P_1$ is positive at every amplitude and detuning, so, to first order in $\beta$, every stable zero crossing of $\mathcal{G}$ shifts outward. Mapping $\mathcal{G}$ over amplitude and detuning gives the attractor diagram \cite{marquardtDynamicalMultistabilityInduced2006}, which Figs.~\ref{fig:energy}(c) and \ref{fig:energy}(d) show for the same cavity at higher input power. Like \cref{fig:energy}(b), these maps assume a sinusoidal trajectory but solve the cavity response exactly, without expanding in $\beta$, so they include $\Delta P_2$ and all higher orders. Without modulation, the stable orbits form a branch that rises with blue detuning and ends near $\bar\Delta\approx3.5\,\Omega_m$, apart from a short detached piece just beyond it. Detuning sweeps steer the resonator along this branch to amplitudes that are not reached from the thermal state at fixed detuning \cite{butersExperimentalExplorationOptomechanical2015}. With full-depth modulation the whole branch moves to larger amplitude, and its end moves out to $\bar\Delta\approx6\,\Omega_m$, raising the largest phase-modulation index from $\mu\approx4.6$ to $7.4$. On the outer part of this branch, however, the optical spring pulls the free-running frequency beyond the locking range of a drive fixed at $\Omega_m$, as the shaded region in \cref{fig:energy}(d) shows. Following the branch there requires the modulation frequency to track the spring shift, by less than $\Gamma_m$ at this operating point. With $\omega_\mathrm{mod}=\Omega_m$ fixed, the highest stable orbit at $\beta=1$ lies at $\mu\approx5.2$. \Cref{fig:energy}(d) also shows a second stable branch at small amplitude for $\bar\Delta\gtrsim3\,\Omega_m$, falling from $\mu\approx1$ to $0.1$. There the added power grows linearly with $A$, as for an external force at $\Omega_m$, so this branch is the resonator's directly driven response to the modulated radiation pressure. It coexists with the self-oscillating branch, making the resonator bistable, and a resonator started from its thermal state settles on it.

Only $\Delta P_1$ is sign-protected: it is linear in the drive sideband and so has no phase-independent part, whereas $\Delta P_2$, quadratic in the sideband, keeps one that is odd in the detuning, as we show in \cref{sup:SecondOrderPower}. This phase-independent part is the dynamical backaction of the drive sidebands themselves, which act as additional pumps at detunings $\bar\Delta\pm\Omega$ carrying a fraction $\beta^2/8$ of the mean power. As the optical damping of each tone is odd in its detuning, their contribution reverses sign between red and blue and vanishes at $\bar\Delta=0$, where equal tones symmetric about resonance exert no backaction damping \cite{clerkBackactionEvasionSqueezing2008, kronwaldArbitrarilyLargeSteadystate2013, malzFloquetApproachBichromatically2016}. The phase-dependent part varies as $2\phi$ instead: the drive tones beat at $2\Omega$ inside the cavity and modulate the optical spring and damping parametrically \cite{steinkeOptomechanicalBackactionevadingMeasurement2013, shomroniTwoToneOptomechanicalInstability2019}, so it too can take either sign. For small $\beta$, $\Delta P_1=\mathcal{O}(\beta)$ dominates $\Delta P_2=\mathcal{O}(\beta^2)$, making the outward shift unconditional. This establishes that the amplitude is enhanced for small modulation depths, but does not guarantee it at stronger modulations. We determine that numerically in \cref{sec:numResults} over the full range $0<\beta\le1$.

\section{Numerical results}
\label{sec:numResults}

To verify and quantify the enhancement that intensity modulation should yield according to \cref{sec:modelEOM}, we perform numerical integration of \cref{eq:mech,eq:cav}.

\subsection{Numerical methods}
\label{subsec:numMethod}

\begin{figure*}[t]
  \centering
  \includegraphics[
    width=\figwidth
  ]{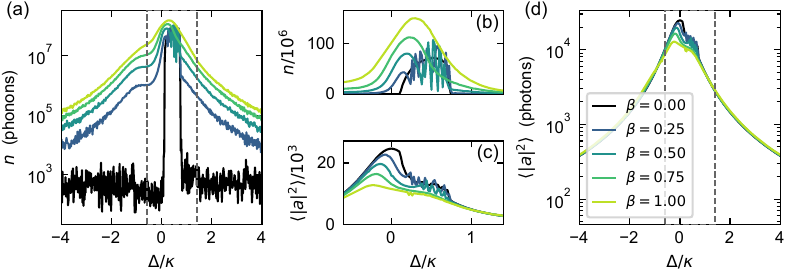}
  \caption{
    \textbf{Intensity modulation raises the peak oscillation amplitude and widens the detuning window beyond the backaction range.}\\
    (a) Phonon occupation $n=\langle x^2\rangle/2x_\mathrm{zpf}^2$ and (d) photon occupation $\langle |a|^2 \rangle$ of the converged periodic steady state versus laser detuning $\Delta/\kappa$, at $C_0=5$ and $\kappa/\Omega_m=2$. Black: without modulation ($\beta=0$); colored: increasing modulation depth. (b),(c) Linear-scale zooms of the dashed regions in (a) and (d), respectively. Backaction-driven self-oscillation occurs only in a narrow blue-detuned window; modulation raises the peak occupation and extends the response beyond that window as a directly driven oscillation.
    \label{fig:singlesweep}
  }
\end{figure*}

We integrate \cref{eq:mech,eq:cav} in dimensionless form, scaling time by $\Omega_m^{-1}$ and displacement by $x_\mathrm{zpf}$. Deterministic trajectories are computed with the adaptive Dormand--Prince scheme \cite{dormandFamilyEmbeddedRungeKutta1980} as implemented in Boost.odeint \cite{ahnertOdeintSolvingOrdinary2011}, with relative and absolute tolerances of $10^{-5}$ and $10^{-7}$. For runs including a thermal bath, the mechanical momentum is subject to a fluctuating force satisfying the fluctuation--dissipation relation at occupation $n_\mathrm{th}$. These runs use a fixed-step Euler--Maruyama scheme \cite{kloedenNumericalSolutionStochastic1992}. In each step, the cavity field is propagated exactly at frozen mechanical position, and the mechanical update uses a semi-implicit Euler rule to preserve long-time energy stability \cite{hairerGeometricNumericalIntegration2006}. Convergence to the periodic steady state is detected by comparing the RMS amplitude and mean intracavity photon number over successive integration windows of duration $2 Q_m/\Omega_m$. When these have converged, the simulation is continued for another 16 periods at fixed step size, and the $x$ and $|a|^2$ traces recorded. From these we extract the steady-state occupations, and the spectra via fast Fourier transform. \Cref{fig:singlesweep} shows the occupations as a function of equilibrium laser detuning and modulation depth.

Including the thermal force serves two purposes in the simulation. First, it provides a physical, parameter-independent seed for the onset of self-oscillation, so the steady state near threshold does not depend on artificial initial displacements. Second, it verifies that the extracted steady-state amplitudes and comb spectra are robust against realistic room-temperature fluctuations ($T=300\,\mathrm{K}$, $n_\mathrm{th}\approx6\times10^{2}$).

Throughout, we use device parameters representative of a GHz-frequency optomechanical crystal. The mechanical mode has frequency $\Omega_m/2\pi=10\,\mathrm{GHz}$, damping $\Gamma_m/2\pi=1\,\mathrm{MHz}$ (quality factor $Q_m=\Omega_m/\Gamma_m=10^4$) and effective mass $m=235\,\mathrm{fg}$ (zero-point motion $x_\mathrm{zpf}=\sqrt{\hbar/2m\Omega_m}\approx1.9\,\mathrm{fm}$). It couples to an optical mode at $\omega_c/2\pi=195\,\mathrm{THz}$ ($\lambda\approx1538\,\mathrm{nm}$) with vacuum optomechanical coupling $g_0/2\pi=1\,\mathrm{MHz}$, and the external coupling is fixed at $\kappa_\mathrm{ex}=\kappa/2$. We sweep the input power $P_0$ and the total cavity linewidth $\kappa$, spanning the sideband-resolved ($\kappa<\Omega_m$) to unresolved ($\kappa>\Omega_m$) regimes. The drive is modulated at the mechanical frequency, $\omega_\mathrm{mod}=\Omega_m$.

We report the sweep in terms of two dimensionless parameters: the sideband-resolution ratio $\kappa/\Omega_m$ and the maximum cooperativity
\begin{equation}
  C_0 \equiv \frac{4 g_0^2\, n_\mathrm{max}}{\kappa\,\Gamma_m}
  = \frac{8 g_0^2\, P_0}{\hbar\omega_L\,\kappa^2\,\Gamma_m},
  \label{eq:C0}
\end{equation}
where $n_\mathrm{max}=2P_0/\hbar\omega_L\kappa$ is the resonant intracavity photon number at critical coupling, $\kappa_\mathrm{ex}=\kappa/2$. Since the photon number realized at finite detuning is smaller, $C_0$ is an upper bound on the cooperativity at each operating point. The sweep spans two decades of $C_0$ for each sideband ratio, defined by using a realistic range of input powers $P_0$ ranging from 10 $\upmu$W to 1 mW, and $\kappa/\Omega_m = 0.1$ to $10$. This means that sideband-resolved devices in the sweep explore a higher range of cooperativities, as they would in a typical experimental scenario.

\subsection{Amplitude enhancement}
\label{subsec:AmpEnhancement}

\begin{figure*}[t]
  \centering
  \includegraphics[
    width=\figwidth
  ]{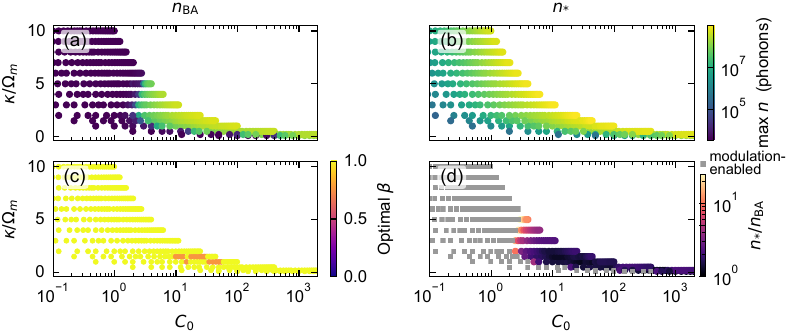}
  \caption{
    \textbf{Intensity modulation enhances the achievable occupation across
    the full parameter space.}\\
    (a) Maximum phonon occupation reached by dynamical backaction alone, $n_\mathrm{BA}$, at each cooperativity $C_0$ and sideband ratio $\kappa/\Omega_m$, optimized over detuning.
    (b) Highest occupation with modulation available, $n_{*}$, optimized over detuning and modulation depth $\beta$, on the same color scale.
    (c) Optimal modulation depth $\beta$, at which the occupation of panel (b) is achieved.
    (d) Enhancement ratio $n_{*}/n_\mathrm{BA}$ of the two maps above. Gray squares mark parameter combinations where backaction alone produces no self-oscillation, so that large-amplitude oscillation there is enabled purely by the modulation. The enhancement is largest near the self-oscillation threshold and decreases to $n_*/n_\mathrm{BA}\approx1.2$--$2$ deep in the self-oscillating regime.
    \label{fig:enhancement}
  }
\end{figure*}

At each combination of cooperativity and sideband ratio, we perform a sweep over detuning and modulation depth, and record the converged mechanical and optical traces. From these we extract the maximum mechanical occupation reached by backaction alone, $n_\mathrm{BA}=\max_\Delta n(\Delta,\beta{=}0)$, and the maximum with modulation available, $n_{*}=\max_{\Delta,\beta} n(\Delta,\beta)$, which Figs.~\ref{fig:enhancement}(a) and \ref{fig:enhancement}(b) show. Because every detuning point is integrated independently from a thermal seed, $n_\mathrm{BA}$ is the amplitude that backaction reaches from the thermal state at a fixed detuning, rather than the larger value that a detuning sweep reaches along the attractor shown in \cref{fig:energy}(c) \cite{marquardtDynamicalMultistabilityInduced2006, butersExperimentalExplorationOptomechanical2015}.

The two maps in Figs.~\ref{fig:enhancement}(a) and \ref{fig:enhancement}(b) differ qualitatively. Without modulation, the occupation remains at the thermal floor until $C_0$ crosses the self-oscillation threshold, where it rises by several orders of magnitude within a single grid step. Since every point starts from a thermal seed, this onset is set by the linear instability $-\Gamma_\mathrm{opt}(\Delta)=\Gamma_m$, and the observed threshold agrees with its minimum over detuning at every sideband ratio to within the grid spacing. The threshold cooperativity is lowest, $C_0\approx2.2$, at $\kappa/\Omega_m\approx2$--$3$. It stays above unity because $C_0$ is defined with the resonant photon number $n_\mathrm{max}$ and because the red sideband partially cancels the gain outside the sideband-resolved regime. Towards the sideband-resolved regime the threshold grows approximately as $4\Omega_m^2/\kappa^2$, because the intracavity photon number at the blue sideband $\Delta=\Omega_m$ is only a fraction of $n_\mathrm{max}$. Towards the sideband-unresolved regime it grows because the sideband asymmetry underlying backaction vanishes. With modulation, there is no threshold: $n_*$ grows continuously from the lowest cooperativity. Below the backaction threshold it scales as $n_*\propto C_0^2$ at each sideband ratio, i.e.\ the amplitude grows linearly with input power, as expected for a directly driven resonator. Above threshold the growth slows to $n_*\propto C_0$, as the oscillation approaches the bandwidth-limited saturation discussed below. At fixed $C_0$, $n_*$ increases with $\kappa/\Omega_m$, since $C_0\propto P_0/\kappa^2$ and a fixed cooperativity at larger linewidth corresponds to a larger input power.

To show that the maximum occupation is always reached by modulating, we plot the modulation depth yielding the maximum occupation in \cref{fig:enhancement}(c), confirming that the maximum is almost always reached at $\beta=1$. At some points in \cref{fig:enhancement}(c) the optimal modulation depth appears below unity. This is because it is the maximum of a quantity that is noisy from run to run, evaluated on a coarse $\beta$ grid. For some parameter combinations, the difference between the $\beta=0.75$ and $\beta=1$ occupations is within the noise level.

\Cref{fig:enhancement}(d) plots the enhancement ratio $n_*/n_\mathrm{BA}$. This magnitude follows directly from the net gain curve's geometry in \cref{fig:energy}(b). Because the modulation enters the energy balance additively, $\mathcal{G}(A_\mathrm{BA},\beta) = \Delta P(A_\mathrm{BA},\beta)$, the enhanced orbit sits at the next downward zero crossing. For moderate $\Delta P$, this outward shift is approximately the added power divided by the slope of the unmodulated gain: $A_*-A_\mathrm{BA} \approx \Delta P(A_\mathrm{BA},\beta)/|\partial_A \mathcal{G}(A_\mathrm{BA},0)|$. This relationship dictates the scaling observed in \cref{fig:enhancement}(d). Near the self-oscillation threshold, the zero crossing has a very low slope, so even a modest $\Delta P$ shifts the orbit significantly. Deep in the self-oscillating regime, the crossing is steep, and a similar added power yields little additional amplitude.

The simulated enhancement follows this trend above threshold. Just above threshold it exceeds an order of magnitude for $\kappa/\Omega_m\geq3$, reaching $n_*/n_\mathrm{BA}\approx10^3$ at $\kappa/\Omega_m=5$. For $\kappa/\Omega_m\leq2$ it decreases with cooperativity to a minimum of 1.2--1.4 at $C_0/C_{0,\mathrm{th}}\approx4$--$7$. At larger cooperativity, however, the enhancement rises again, to 1.6--2.3 at the highest cooperativities in the sweep, rather than approaching unity. This regime lies outside the perturbative picture above, which compares both orbits at the same detuning, since here the optimal detuning with modulation separates from that of backaction alone. In the sideband-resolved regime, $\kappa/\Omega_m\leq0.5$, backaction remains optimal at $\Delta\approx\Omega_m$, while the modulated optimum jumps to $\Delta\approx2\Omega_m$. There, the lower modulation sideband of the drive lies at the blue mechanical sideband of the cavity. In the sideband-unresolved regime, backaction is optimal at the blue edge of its self-oscillation window, which moves outward with cooperativity, while the modulated optimum remains closer to resonance. For $\kappa/\Omega_m\geq4$, the sweep ends before this regime is reached.

The decrease above threshold also has a more intuitive origin. At large mechanical amplitudes, the mechanics strongly modulates the optics, $\mu \gg 1$, distributing the optical field across about $2\mu$ sidebands, of which only about $\kappa/\Omega_m$ fall within the cavity line. Their remaining contributions to $P_\mathrm{rp}$ largely cancel, leading to a saturation set by the finite cavity bandwidth \cite{pootBackactionLimitsSelfsustained2012}. This is the same limit that bounds the far-blue tip of the backaction attractor in \cref{fig:energy}(c), which detuning sweeps approach along the contour \cite{butersExperimentalExplorationOptomechanical2015}. Modulation adds a bounded $\Delta P$ to the energy balance and shifts the tip outward, as \cref{fig:energy}(d) shows, but cannot bypass the bandwidth limit itself. Deep in the self-oscillating regime, static backaction already drives the resonator close to this saturation scale at fixed detuning, so the added power shifts the orbit only slightly. The remaining enhancement at the highest cooperativities stems from the modulated optimum moving to a different detuning. Consequently, modulation is most advantageous for reaching the saturation amplitude under conditions where standard backaction at fixed detuning is insufficient.

We additionally observe that, at equal distance from threshold, the enhancement grows with the sideband ratio once $\kappa/\Omega_m\gtrsim1$. At $C_0/C_{0,\mathrm{th}}=2$ it increases from $n_*/n_\mathrm{BA}\approx1.4$ in the sideband-resolved regime to $4.5$ at $\kappa/\Omega_m=4$. For $\kappa/\Omega_m<1$ it is nearly independent of the sideband ratio. This is because the unmodulated backaction gain relies on the asymmetric cavity response between the two motional sidebands, which vanishes as the cavity response flattens ($\kappa \gg \Omega_m$). In contrast, the first-order modulated power $\Delta P_1$ stems from the beat between the mean field and the drive sidebands. This survives a flat cavity response because $K(A)$ depends on the magnitudes of the susceptibility products, not their differences. In the region with gray squares in \cref{fig:enhancement}(d), backaction alone never reaches threshold, making modulation the sole driver of large-amplitude motion. Here, the system acts as a driven oscillator rather than a self-sustained one: the oscillation emerges without a Hopf bifurcation and the amplitude is continuous with drive strength, similar to the threshold-free behavior seen in two-tone driving \cite{guOpticalFrequencyComb2024}.

Intensity-modulated driving is therefore most valuable for sideband-unresolved systems and under constrained power budgets. These are the operating conditions typical of integrated platforms, where absorption-induced thermal effects limit the tolerable intracavity power: circulating light heats the cavity and drags the resonance, producing thermal bistability and drift \cite{carmonDynamicalThermalBehavior2004, notomiOpticalBistableSwitching2005}. Our equal-power comparison is fair under such a thermal constraint. The instantaneous input power reaches $2P_0$ at full modulation depth, but the thermal load is governed by the cycle-averaged absorbed power, which remains fixed.

\subsection{Frequency comb scaling}
\label{subsec:combscaling}

As the mechanical oscillation phase-modulates the optical field with phase-modulation index $\mu$, the optical field picks up a series of sidebands at harmonics with frequencies $n \cdot \Omega$ and height $J_n(\mu)$, forming an optical frequency comb \cite{marquardtDynamicalMultistabilityInduced2006, miriOptomechanicalFrequencyCombs2018, mercadeMicrowaveOscillatorFrequency2018, huGenerationOpticalFrequency2021, gouChipScaleOptomechanicalFrequency2025}. When the amplitude $A$ of the mechanical oscillation is increased, this effectively increases the modulation depth and thereby the width of the comb generated by the optomechanical system. Although a pure phase modulation would leave the intensity constant and the comb invisible to a photodiode, the detuned cavity susceptibility weights the sidebands unequally in amplitude and phase. The comb therefore also appears in the intensity spectrum, as a microwave frequency comb \cite{huGenerationOpticalFrequency2021} whose width should likewise grow with the mechanical amplitude.

\begin{figure*}[t]
  \centering
  \includegraphics[
    width=\figwidth
  ]{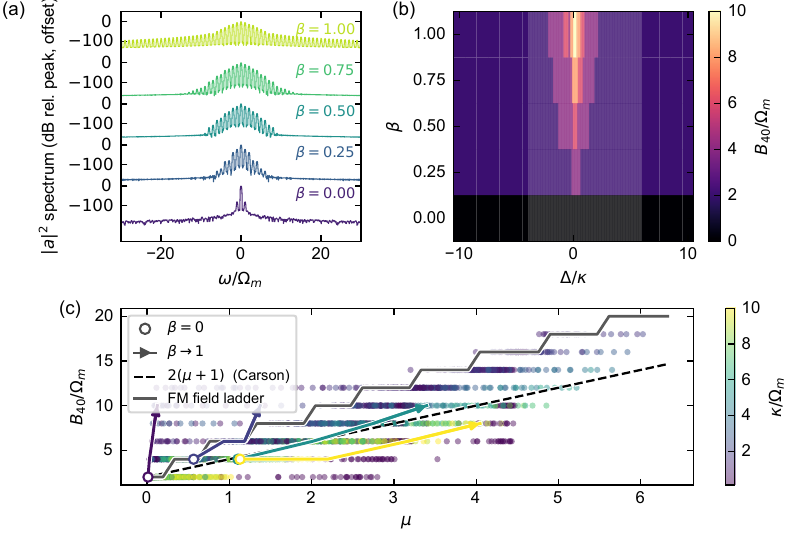}
  \caption{
    \textbf{The comb bandwidth increases with the phase-modulation index, which is deepened by the modulation-enhanced amplitude.}\\
    (a) Spectra of the intracavity photon number $|a|^2(t)$ in the converged periodic steady state at increasing modulation depth $\beta$, at $P_0 = 500\,\upmu\mathrm{W}$ and $\kappa/\Omega_m = 5$ ($C_0 \approx 2$). Each spectrum is shown at its bandwidth-maximizing detuning, in dB relative to its peak and offset vertically. The comb widens monotonically with $\beta$.
    (b) Contiguous $40\,\mathrm{dB}$ comb bandwidth $B_{40}$, in units of $\Omega_m$, versus detuning $\Delta/\kappa$ and modulation depth $\beta$ for the same device; modulation broadens both the comb and the detuning window in which one forms.
    (c) Maximum $B_{40}$ versus phase-modulation index $\mu = GA/\Omega$ for every $(C_0, \kappa/\Omega_m, \beta)$ combination of the sweep, colored by $\kappa/\Omega_m$. Trajectories trace $\beta = 0$ (open circle) to $\beta = 1$ (arrowhead) at fixed $C_0 \approx 2$ for $\kappa/\Omega_m = 0.5$, $2$, $5$, and $C_0 \approx 1$ for $\kappa/\Omega_m=10$. Modulation moves each device to larger $\mu$ and hence wider combs. Dashed line: Carson's rule $B \approx 2(\mu+1)\,\Omega_m$ \cite{carsonNotesTheoryModulation1922}; solid staircase: the parameter-free frequency modulation field ladder, obtained by applying the same $-40\,\mathrm{dB}$ walk to the bare Bessel sideband amplitudes $|J_n(\mu)|$.
    \label{fig:combscaling}
  }
\end{figure*}

To evaluate this, we calculate the spectra of the traces at the end of the simulation, as defined in \cref{subsec:numMethod}. An example of a collection of intensity spectra, in this case for various modulation depths at $P_0=500\,\upmu$W and $\kappa/\Omega_m=5$ ($C_0\approx2$), is given in \cref{fig:combscaling}(a). We see an increase in the comb width as we increase the modulation depth and thereby the mechanical oscillation amplitude.

To quantify this effect we evaluate the contiguous comb bandwidth $B_{D}$. Denoting by $P_k = \bigl|\mathcal{F}\{|a|^2\}(k\,\omega_\mathrm{mod})\bigr|^2$ the power of the comb tooth at the $k$-th harmonic of the modulation frequency, we define the contiguous span
\begin{equation}
  B_D = \left(k_+ - k_-\right) \omega_\mathrm{mod},
  \label{eq:BD}
\end{equation}
where $k_\pm$ are the outermost harmonics reachable from the strongest tooth through consecutive teeth. All of these teeth must remain within $D$ decibels of the strongest one: $10\log_{10}\!\left(P_k/\max_j P_j\right) \ge -D$. Isolated teeth beyond the first sub-threshold gap do not contribute, as they do not form a usable equidistant comb \cite{herrUniversalFormationDynamics2012}. The threshold $D$ is a convention: comb and supercontinuum spans are quoted anywhere from 10 to 60~dB below the peak \cite{torres-companyOpticalFrequencyComb2014,huangLowphasenoise18GHz2015,caoSupercontinuumGenerationNonlinear2022}. We take $D=40$~dB, a common choice for supercontinuum spans \cite{ciretGenerationUltrabroadbandCoherent2017,caoSupercontinuumGenerationNonlinear2022}. It corresponds to the 1\% amplitude criterion used to count significant FM sidebands \cite{haykinCommunicationSystems2010}, and to the typical dynamic range of optical spectrum measurements \cite{zhangBroadbandElectroopticFrequency2019}.

Evaluating $B_{40}$ at every modulation depth and detuning for the parameters of \cref{fig:combscaling}(a) and normalizing by the mechanical frequency $\Omega_m$ yields \cref{fig:combscaling}(b). The combs with maximum span for a given modulation depth are the examples displayed in \cref{fig:combscaling}(a). From \cref{fig:combscaling}(b) we can observe that not only does the comb bandwidth increase with increased $\beta$, so does the detuning window in which a comb is formed.

We extend this analysis to all $(C_0,\kappa/\Omega_m)$ combinations of the sweep. \Cref{fig:combscaling}(c) plots the maximum $B_{40}$ for each value of $\beta$ in each combination against the phase-modulation index $\mu$ of \cref{eq:mu}. Each point is one $(C_0,\kappa/\Omega_m,\beta)$ combination, colored by its sideband ratio, at the largest-amplitude detuning that maximizes $B_{40}$. We use $\mu$ rather than $C_0$ or $\beta$ on the horizontal axis because, through the Bessel weights in \cref{eq:a0}, it sets the width of the sideband ladder. This allows devices with very different input powers and linewidths to be compared on the same axis. Since $|a|^2$ is real, its spectrum is symmetric about the carrier, and $B_{40}$ only takes even multiples of $\Omega_m$. The points therefore fall on horizontal rows, each row adding one pair of comb teeth to the one below.

Two trends are visible in \cref{fig:combscaling}(c). First, $B_{40}$ increases with $\mu$ over the full sweep: devices that oscillate at larger amplitude generate wider combs, regardless of whether that amplitude is reached by backaction or by modulation. Second, the comb width at fixed $\mu$ depends on the sideband ratio. It is largest for $\kappa/\Omega_m\approx0.5$--$2$ and decreases in both directions. At $\mu\approx4$--$5$, $B_{40}$ drops from $14$--$16\,\Omega_m$ at $\kappa/\Omega_m\approx1$ to $8\,\Omega_m$ at $\kappa/\Omega_m=10$, and to $6\,\Omega_m$ at $\kappa/\Omega_m=0.1$. This follows from how the intensity comb is formed. The mechanical motion only modulates the phase of the intracavity field, and a pure phase modulation leaves $|a|^2$ constant: for a flat cavity response the sum in \cref{eq:a0} reduces to $e^{-i\mu\sin\Omega t}$, cancelling the prefactor. An intensity comb therefore only appears because the cavity susceptibility weights the sidebands at $\bar\Delta+n\Omega$ unequally, so that the beat notes between them no longer cancel. This phase-to-amplitude conversion is weak in two limits. For $\kappa\gg\Omega_m$ the cavity response is nearly flat over the sidebands carrying significant weight, and the outer teeth fall below the $-40\,$dB threshold at lower harmonic order. For $\kappa\ll\Omega_m$ the cavity passes essentially only the sideband closest to resonance, so the intracavity field approaches a single tone and the beat notes that form the intensity comb are suppressed. The conversion is most efficient for $\kappa\sim\Omega_m$, where the susceptibility varies strongly between neighboring sidebands while still passing several of them.

The expected dependence of the bandwidth on $\mu$ can be reasoned from Carson's rule \cite{carsonNotesTheoryModulation1922}, stating that the bandwidth should grow linearly with the modulation depth: $B\approx2(\mu+1)\Omega_m$. This is indicated by the dashed line in \cref{fig:combscaling}(c). To show the influence of $\beta$ more explicitly, we draw arrows connecting $\beta=0$ to $\beta=1$ for $\kappa/\Omega_m=0.5$, 2.0, 5.0, and 10.0, all at $C_0=2.0$. The exception is $\kappa/\Omega_m=10$, where this value is not reached within our sweep, and we plot $C_0=1.0$ instead. As $\beta$ increases, the arrows move to higher $\mu$ due to the increased mechanical amplitude, and consequently to higher bandwidths. Adding intensity modulation to an optomechanical system thus enhances its performance for generating optical frequency combs.

Carson's rule captures the leading trend but not all the measured spans at a 40 dB threshold, which reach into the roll-off of the sideband ladder beyond the Carson edge. As an additional reference, we compute the span of an ideal frequency-modulated field directly: applying the same contiguous 40 dB walk of \cref{eq:BD} to the bare Jacobi--Anger sideband amplitudes $|J_n(\mu)|$ yields the solid staircase in \cref{fig:combscaling}(c). This curve does not depend on any system parameters --- its only input is $\mu$ itself --- and represents the comb width that the mechanical phase modulation imprints on the field before any cavity filtering. It exceeds Carson's estimate by the width of the Bessel transition zone: near the ladder edge $|k|\approx\mu$ the sideband amplitudes roll off as an Airy function over $\sim\mu^{1/3}$ harmonics \cite{NIST:DLMF:10.19iii}. At small $\mu$ and large $\beta$, points lie above the pure frequency modulation ladder for a different reason. The drive carries harmonics of its own, which set a floor on the span independent of the mechanical motion.

Taken together, \cref{fig:combscaling}(c) shows that the comb bandwidth is set primarily by the mechanical amplitude through $\mu$ \cite{huGenerationOpticalFrequency2021}, so the amplitude enhancement of \cref{subsec:AmpEnhancement} carries over directly into wider combs at equal mean input power. The cavity then determines how much of the phase-modulation ladder appears in the intensity spectrum. For $\kappa\sim\Omega_m$ the measured spans follow the ideal frequency modulation ladder, while for strongly sideband-resolved or unresolved cavities they fall below it. This creates a trade-off in the sideband-unresolved regime: there, modulation yields the largest gain in $\mu$, but a smaller fraction of it is converted into comb teeth, as the $\kappa/\Omega_m=10$ trajectory illustrates. Consequently, at every input power in the sweep, the widest comb at $\beta=1$ is obtained for $\kappa/\Omega_m\approx0.5$--$2$, making intensity-modulated devices with $\kappa\sim\Omega_m$ the preferred choice for broadband comb generation under a fixed power budget.

\section{Conclusion}
We have carried out a systematic comparison of backaction-only and intensity-modulated driving of an optomechanical resonator, each evaluated at its own optimal operating point, at fixed mean input power and cavity linewidth. A cycle-averaged energy balance shows that the modulation enters the net gain additively, and that its first-order contribution $\Delta P_1=\beta K(A)\cos\psi$ is sign-protected. Being linear in the drive sideband, it is a zero-mean sinusoid in the modulation phase for every amplitude and detuning. Injection locking settles the drive on the energy-injecting phase wherever the drive locks robustly. Wherever the drive stays locked near resonance, the stable periodic orbit is therefore guaranteed to shift outward, $A_*(\beta)>A_\mathrm{BA}$, at equal mean power. At a given average laser power, modulation thus increases the mechanical amplitude that can be reached.

Numerical integration of the full equations of motion confirms this enhancement across the entire swept parameter space and quantifies it well beyond the perturbative regime, with the largest amplitudes reached near full modulation depth. The enhancement is greatest near the self-oscillation threshold and in the sideband-unresolved regime, where backaction gain relies on a susceptibility contrast that a broad cavity response suppresses, while the directly injected power survives. For the realistic powers used in this work, backaction alone never reaches threshold in this regime, and the modulation turns the system into a threshold-free driven oscillator. Deep in the self-oscillating regime the enhancement decreases to $n_*/n_\mathrm{BA}\approx1.2$--$2$, as both driving schemes approach the same saturation set by the finite cavity bandwidth. Because the mechanical amplitude sets the optical phase-modulation index $\mu$, the amplitude enhancement translates directly into wider frequency combs, as the contiguous comb bandwidth increases with $\mu$.

These results identify intensity-modulated driving as a practical route to larger oscillation amplitudes and wider combs, precisely under the conditions typical of integrated platforms. There, mean input powers are capped by absorption-induced thermal effects, a constraint that the modulation respects since the cycle-averaged power is unchanged. Full-depth intensity modulation at GHz frequencies is readily available with commercial electro-optic modulators, and the scheme is open loop, avoiding the detection and processing overhead of feedback-based enhancement. As the oscillation frequency is inherited from the external RF source, the comb spacing is moreover locked against optical spring drift.

Our comparison concerns a resonator switched on at fixed detuning, but since, with the drive locked, the first-order modulated power is positive at every amplitude and detuning, the modulation deforms the entire attractor diagram \cite{marquardtDynamicalMultistabilityInduced2006}. We expect this deformation to be strongest near the folds where stable and unstable branches merge: there the slope of the net gain vanishes, so that, as near the self-oscillation threshold, a modest added power moves the zero crossing substantially. \Cref{fig:energy}(d) shows that the modulation extends the branch followed in detuning sweeps \cite{butersExperimentalExplorationOptomechanical2015} to larger blue detuning and amplitude, provided the drive frequency tracks the optical spring shift to stay within the locking range. By lifting the net gain above zero across the gap that separates high-amplitude branches from the lower ones, it could also make these branches accessible directly from the thermal state, where they are otherwise reachable only along a prepared path. How modulation depth and phase reshape and connect the coexisting attractors \cite{krauseNonlinearRadiationPressure2015}, and whether they permit deterministic switching between them, is a natural subject for further study, alongside extensions to multimode systems and combined modulation and feedback strategies.

\begin{acknowledgments}
  We thank Juha Muhonen and Jorien van der Meulen for useful discussions. The C++ simulation code and the Python scripts for data analysis and plotting were developed and debugged with the help of Claude Code (Anthropic), using the Claude Opus 4.8 and 5 models. The authors verified all code and numerical results; the physical interpretation is their own. This work is part of the research programme of The Netherlands Organisation for Scientific Research (NWO). It is supported by an NWO Vrij Programma (680-92-18-04) grant. It is also funded by the European Union (EquSpace, Grant Agreement No. 101185817 and Q-MEME, Grant Agreement No. 101088055). Views and opinions expressed are however those of the authors only and do not necessarily reflect those of the European Union or the European Research Council Executive Agency. Neither the European Union nor the granting authority can be held responsible for them.
\end{acknowledgments}

\section*{Data availability}
Data underlying the results presented in this paper, together with the simulation and plotting code, are available in Ref.~\cite{jansenZenodo2026}.

\appendix

\section{First-order modulation power}
\label{sup:FirstOrderPower}

To evaluate $\Delta P_1$ we first need the cavity field on the prescribed trajectory $x(t)=\bar x+A\cos\Omega t$. Substituting it into \cref{eq:cav} splits the position-dependent detuning into a static and an oscillating part,

\begin{equation}
  \dot a=\bigl(i\bar\Delta-\tfrac\kappa2\bigr)a+iGA\cos(\Omega t)\,a+\sqrt{\kappa_\mathrm{ex}}\,a_\mathrm{in},
  \label{eq:cav-traj}
\end{equation}
where $\bar\Delta = \Delta + G \bar x$ is the detuning renormalized by the static displacement. The oscillating term only rotates the phase of $a$, so we move to a frame that removes it, the classical counterpart of the polaron transformation that shifts the mechanical dependence into the phase of the drive \cite{rablPhotonBlockadeEffect2011, nunnenkampSinglePhotonOptomechanics2011, lorchLaserTheoryOptomechanics2014}. Writing $a=e^{i\theta}c$ with $\theta=\mu\sin\Omega t$, the phase derivative $\dot\theta=GA\cos\Omega t$ cancels the oscillating term exactly and leaves a constant-coefficient equation for $c$,

\begin{equation}
  \dot c=\bigl(i\bar\Delta-\tfrac\kappa2\bigr)c+\sqrt{\kappa_\mathrm{ex}}\,e^{-i\theta}a_\mathrm{in}.
  \label{eq:ceq}
\end{equation}

The mechanical motion now enters only as a phase modulation of the drive, with the depth $\mu$ of \cref{eq:mu}, which also sets the comb width in \cref{subsec:combscaling}. The Jacobi--Anger expansion \cite{NIST:DLMF:10.12ii}, $e^{-i\mu\sin\Omega t}=\sum_n J_n(\mu)e^{-in\Omega t}$, decomposes this modulated drive into sidebands at multiples of $\Omega$. Because $e^{-i\theta}$ now appears only in the source term, \cref{eq:ceq} is linear with a constant coefficient on $c$, and each sideband can be solved separately. Inserting $c\propto e^{-in\Omega t}$ shows that, once transients have decayed at rate $\kappa/2$, a source at $e^{-in\Omega t}$ drives a field weighted by the cavity susceptibility

\begin{equation}
  \chi_n\equiv\frac{1}{\tfrac\kappa2-i(\bar\Delta+n\Omega)}.
  \label{eq:chi}
\end{equation}

The mean field ($a_\mathrm{in}=\bar a_\mathrm{in}$) is then $c^{(0)}=\sum_n u_n e^{-in\Omega t}$ with $u_n=\sqrt{\kappa_\mathrm{ex}}\,\bar a_\mathrm{in}J_n\chi_n$, which after undoing the phase transformation, $a^{(0)}=e^{i\theta}c^{(0)}$, is \cref{eq:a0}. Each Bessel sideband of the phase-modulated drive is thus filtered by the cavity Lorentzian at detuning $\bar\Delta+n\Omega$ \cite{marquardtDynamicalMultistabilityInduced2006, krauseNonlinearRadiationPressure2015, huGenerationOpticalFrequency2021}. At resonance $\omega_\mathrm{mod}=\Omega$ with control phase $\phi$, the first-order drive sideband $\delta a_\mathrm{in}=\tfrac\beta2\bar a_\mathrm{in}\cos(\Omega t-\phi)$ enters \cref{eq:ceq} multiplied by $e^{-i\theta}$. Writing the cosine as two exponentials shifts each Bessel term $J_n e^{-in\Omega t}$ to $e^{-i(n\pm1)\Omega t}$, so after relabelling $k=n\pm1$ the source at $e^{-ik\Omega t}$ carries $J_{k\mp1}$. Weighting each by $\chi_k$ gives $\delta c=\sum_k v_k e^{-ik\Omega t}$ with

\begin{equation}
  v_k=\sqrt{\kappa_\mathrm{ex}}\,\bar a_\mathrm{in}\,\tfrac\beta4\,\chi_k\bigl(J_{k+1}e^{-i\phi}+J_{k-1}e^{+i\phi}\bigr),
  \label{eq:v}
\end{equation}
where $J_n\equiv J_n(\mu)$ are real.

The cross-term force is $\delta F=2\hbar G\,\mathrm{Re}[a^{(0)}{}^{\!*}\delta a]$. The phase factors cancel, $a^{(0)}{}^{\!*}\delta a=c^{(0)}{}^{\!*}\delta c$, so its Fourier components $\delta F=\sum_p\hat F_p e^{-ip\Omega t}$ are

\begin{equation}
  \hat F_p=\hbar G\sum_n\bigl(u_n^*v_{n+p}+u_n v_{n-p}^*\bigr),\qquad \hat F_{-p}=\hat F_p^{*}.
  \label{eq:Fhat}
\end{equation}

This is the convolution of sideband amplitudes that gives the unmodulated radiation-pressure force, $\hat F_p^{(0)}=\hbar G\sum_n u_n^*u_{n+p}$ \cite{marquardtDynamicalMultistabilityInduced2006, ludwigOptomechanicalInstabilityQuantum2008}, now taken between the mean-field comb $u_n$ and the drive-sideband comb $v_k$. Only the velocity quadrature does work. With $\dot x=-A\Omega\sin\Omega t$ and $\langle f\sin\Omega t\rangle=\mathrm{Im}\hat f_1$,

\begin{equation}
  \Delta P_1=\langle\delta F\,\dot x\rangle=-A\Omega\,\mathrm{Im}\hat F_1.
  \label{eq:dP1sup}
\end{equation}

The same projection applied to $\hat F_1^{(0)}$ gives the backaction power $P_\mathrm{rp}^{(0)}=-A\Omega\,\mathrm{Im}\hat F_1^{(0)}$, the Bessel-sum form of the attractor diagram \cite{marquardtDynamicalMultistabilityInduced2006}.

Evaluating \cref{eq:Fhat} at $p=1$ gives

\begin{equation}
  \hat F_1=\hbar G\sum_n\bigl(u_n^*\,v_{n+1}+u_n\,v_{n-1}^*\bigr).
\end{equation}

Substituting $u_n$ and $v_k$, every term carries $\hbar G\,\kappa_\mathrm{ex}|\bar a_\mathrm{in}|^2\equiv\Lambda$ (real), and the shifted Bessel labels are

\begin{align}
  v_{n+1}   &=\sqrt{\kappa_\mathrm{ex}}\,\bar a_\mathrm{in}\,\tfrac\beta4\,\chi_{n+1}\bigl(J_{n+2}e^{-i\phi}+J_{n}e^{+i\phi}\bigr),\\
  v_{n-1}^* &=\sqrt{\kappa_\mathrm{ex}}\,\bar a_\mathrm{in}^*\,\tfrac\beta4\,\chi_{n-1}^*\bigl(J_{n}e^{+i\phi}+J_{n-2}e^{-i\phi}\bigr).
\end{align}

The two products become (using $J_n\in\mathbb R$)

\begin{align}
  \hbar G\,u_n^*\,v_{n+1}   &=\frac{\Lambda\beta}{4}\,J_n\,\chi_n^*\chi_{n+1}\bigl(J_{n+2}e^{-i\phi}+J_{n}e^{+i\phi}\bigr),\\
  \hbar G\,u_n\,v_{n-1}^*   &=\frac{\Lambda\beta}{4}\,J_n\,\chi_n\chi_{n-1}^*\bigl(J_{n}e^{+i\phi}+J_{n-2}e^{-i\phi}\bigr).
\end{align}

Summing over $n$ and collecting the two phase factors,

\begin{equation}
  \begin{split}
  \hat F_1=\frac{\Lambda\beta}{4}\Biggl[&
    e^{+i\phi}\sum_n J_n^2\bigl(\chi_n^*\chi_{n+1}+\chi_n\chi_{n-1}^*\bigr)\\
  &+e^{-i\phi}\sum_n J_n\bigl(\chi_n^*\chi_{n+1}J_{n+2}+\chi_n\chi_{n-1}^*J_{n-2}\bigr)\Biggr],
  \end{split}
  \label{eq:F1}
\end{equation}
which is $\hat F_1=\beta(\Sigma_+e^{+i\phi}+\Sigma_-e^{-i\phi})$ with $\Sigma_\pm$:

\begin{equation}
  \begin{split}
  \Sigma_+&=\frac{\Lambda}{4}\sum_n J_n^2\bigl(\chi_n^*\chi_{n+1}+\chi_n\chi_{n-1}^*\bigr),\\
  \Sigma_-&=\frac{\Lambda}{4}\sum_n J_n\bigl(\chi_n^*\chi_{n+1}J_{n+2}+\chi_n\chi_{n-1}^*J_{n-2}\bigr).
  \end{split}
  \label{eq:Spm}
\end{equation}

In the unmodulated force harmonic $\hat F_1^{(0)}$ the susceptibility products $\chi_n^*\chi_{n+1}$ carry the pair $J_nJ_{n+1}$ \cite{lorchLaserTheoryOptomechanics2014, russoAnalyticalExplorationOptomechanical2025}. The drive sideband shifts one Bessel index by $\pm1$, producing the pairs $J_n^2$ and $J_nJ_{n\pm2}$ of $\Sigma_\pm$. The same Bessel pairs on the susceptibility product $\chi_n\chi_{n-1}^*$ appear in the phase-drift coefficient of Ref.~\cite{amitaiSynchronizationOptomechanicalSystem2017}, which is the phase quadrature of this force harmonic, whereas here we keep its velocity quadrature. This makes $\mathrm{Im}\hat F_1=\beta\bigl[\mathrm{Re}(\Sigma_+-\Sigma_-)\sin\phi+\mathrm{Im}(\Sigma_++\Sigma_-)\cos\phi\bigr]$ a single sinusoid in $\phi$, which we write as $R\cos(\phi-\phi_0)$ with amplitude $R=\sqrt{\mathrm{Re}(\Sigma_+-\Sigma_-)^2+\mathrm{Im}(\Sigma_++\Sigma_-)^2}$ and reference phase

\begin{equation}
  \tan\phi_0(A)=\frac{\mathrm{Re}(\Sigma_+-\Sigma_-)}{\mathrm{Im}(\Sigma_++\Sigma_-)},
  \label{eq:phi0}
\end{equation}
the modulation phase at which the injected first-order power is extremal. \Cref{eq:dP1sup} then reduces to the main-text form $\Delta P_1=\beta K(A)\cos\psi$ with $\psi\equiv\phi-\phi_0(A)+\pi$ and $K(A)=A\Omega R$ as in \cref{eq:K-paper}. The sign protection follows from $\Delta P_1$ being an interference term: the beat of the mean field with the field driven by the first-order sideband, linear in $\delta a_\mathrm{in}\propto\cos(\Omega t-\phi)$. The phase therefore enters only through $e^{\pm i\phi}$, as \cref{eq:F1} shows, and $\Delta P_1$ is a zero-mean sinusoid in $\phi$. Shifting the phase by $\pi$ flips the sideband and turns constructive into destructive interference. At every amplitude and detuning, half of all phases thus inject power, at most $\Delta P_1=\beta K$ at $\psi=0$. Injection locking settles on this branch, as shown in \cref{supp:locking}. $\Delta P_2$, built from $|\delta a^{(1)}|^2$ and the $\cos^2(\Omega t-\phi)$ drive correction, is instead blind to the sign of the sideband. As \cref{sup:SecondOrderPower} shows, it depends on the phase only through $e^{\pm2i\phi}$, so it is unchanged under $\phi\to\phi+\pi$, and it is odd in the detuning, reversing sign between red and blue.

\section{Second-order power}
\label{sup:SecondOrderPower}

$\Delta P_2$ collects the two $\mathcal{O}(\beta^{2})$ contributions to \cref{eq:Prp-orders}: the first-order sideband intensity $\hbar G\,|\delta a^{(1)}|^{2}$, and the beat $2\hbar G\,\mathrm{Re}[a^{(0)}{}^{\!*}\delta a^{(2)}]$ of the mean field with the second-order drive correction $\delta a_\mathrm{in}^{(2)}=-\tfrac{\beta^{2}}{8}\bar a_\mathrm{in}\cos^{2}(\Omega t-\phi)$. Both contribute to the fundamental of the second-order force, and like the first order only its velocity quadrature does work, so $\Delta P_2=-A\Omega\,\mathrm{Im}\,\hat F_1^{(2)}$ with $\hat F_1^{(2)}=\hat F_{1,i}^{(2)}+\hat F_{1,ii}^{(2)}$.

The intensity term follows from $|\delta a^{(1)}|^{2}=|\delta c|^{2}$ and the sideband amplitudes $v_k$ of \cref{eq:v},
\begin{equation}
  \hat F_{1,i}^{(2)} = \hbar G\sum_k v_k\,v_{k-1}^{*}.
  \label{eq:F1i}
\end{equation}
For the beat term, the second-order drive enters \cref{eq:ceq} through $\sqrt{\kappa_\mathrm{ex}}\,e^{-i\theta}\delta a_\mathrm{in}^{(2)}$. Expanding $\cos^{2}(\Omega t-\phi)$ into a static part and components at $\pm2\Omega$ and applying the Jacobi--Anger expansion as before gives $\delta c^{(2)}=\sum_k w_k e^{-ik\Omega t}$ with
\begin{equation}
  w_k = -\frac{\beta^{2}}{32}\,\sqrt{\kappa_\mathrm{ex}}\,\bar a_\mathrm{in}\,
  \chi_k\bigl(2J_k + J_{k-2}\,e^{+2i\phi} + J_{k+2}\,e^{-2i\phi}\bigr),
  \label{eq:w}
\end{equation}
and beating against the mean field $c^{(0)}=\sum_n u_n e^{-in\Omega t}$,
\begin{equation}
  \hat F_{1,ii}^{(2)} = \hbar G\sum_n\bigl(u_n^{*}\,w_{n+1} + u_n\,w_{n-1}^{*}\bigr).
  \label{eq:F1ii}
\end{equation}

Where the first-order force of \cref{eq:F1} carried a single modulation harmonic $e^{\pm i\phi}$, squaring the drive gives the second-order force a static part and a second harmonic $e^{\pm2i\phi}$. Collecting \cref{eq:F1i,eq:F1ii} onto the common basis $\chi_k\chi_{k-1}^{*}$ yields,
\begin{equation}
  \hat F_1^{(2)} = \beta^{2}\bigl(\Pi_0 + \Pi_+\,e^{+2i\phi} + \Pi_-\,e^{-2i\phi}\bigr),
\end{equation}
with $J_n\equiv J_n(\mu)$, and
\begin{align}
  \Pi_0 &= \frac{\Lambda}{16}\sum_k \chi_k\chi_{k-1}^{*}
  \bigl(J_{k+1}J_k + J_{k-1}J_{k-2} - 2J_kJ_{k-1}\bigr), \label{eq:Pi0}\\
  \Pi_+ &= \frac{\Lambda}{32}\sum_k \chi_k\chi_{k-1}^{*}
  \bigl(2J_kJ_{k-1} - J_{k+1}J_k - J_{k-1}J_{k-2}\bigr) \nonumber\\
  &= -\tfrac12\,\Pi_0, \label{eq:Pip}\\
  \Pi_- &= \frac{\Lambda}{32}\sum_k \chi_k\chi_{k-1}^{*}
  \bigl(2J_{k+1}J_{k-2} - J_{k-1}J_{k+2} - J_kJ_{k-3}\bigr). \label{eq:Pim}
\end{align}
The relation $\Pi_+=-\tfrac12\Pi_0$ follows by relabelling the beat sum onto the $\chi_k\chi_{k-1}^{*}$ index of the intensity sum, so that
\begin{equation}
  \Delta P_2 = -A\Omega\,\beta^{2}\,
  \mathrm{Im}\!\left[\Pi_0\bigl(1-\tfrac12 e^{+2i\phi}\bigr) + \Pi_-\,e^{-2i\phi}\right].
  \label{eq:dP2full}
\end{equation}

$\Delta P_1$ is a zero-mean sinusoid in $\phi$ with amplitude $\beta K\ge0$ for every $A$ and detuning. $\Delta P_2$, in contrast, has a phase-independent part, and each $\Pi$ is the imaginary part of a sum over products of distinct susceptibilities $\chi_k\chi_{k-1}^{*}$, whose sign is not fixed. The phase-independent part is carried entirely by the static term, $\Delta P_2\big|_{\phi\text{-indep}}=-A\Omega\,\beta^{2}\,\mathrm{Im}\,\Pi_0$. Using $\chi_n(-\bar\Delta)=\chi_{-n}^{*}(\bar\Delta)$ and $J_{-n}=(-1)^{n}J_n$, the relabelling $k \mapsto 1-k$ maps the susceptibility product $\chi_k\chi_{k-1}^{*}$ onto itself while flipping the sign of every Bessel factor. This means that each coefficient is odd in the mean detuning, $\Pi_0(-\bar\Delta)=-\Pi_0(\bar\Delta)$ and $\Pi_\pm(-\bar\Delta)=-\Pi_\pm(\bar\Delta)$. The latter is also consistent with the exact identity $\Pi_+=-\tfrac{1}{2}\Pi_0$. Since $\Delta P_2$ is a fixed linear combination of these coefficients at fixed control phase $\phi$, it inherits the same parity,

\begin{equation}
  \Delta P_2(-\bar\Delta,\phi) = -\,\Delta P_2(\bar\Delta,\phi),
  \label{eq:dP2parity}
\end{equation}
vanishing at $\bar\Delta=0$ and reversing between red and blue detuning. The first-order term of \cref{eq:dP1-paper} lacks this sign freedom. This is why the outward shift is guaranteed only in the perturbative interior of the band, where $\Delta P_1=\mathcal{O}(\beta)$ dominates $\Delta P_2=\mathcal{O}(\beta^{2})$. At full modulation depth $\Delta P_2$ could become comparable to $\Delta P_1$, and its contribution is what the numerical sweep of \cref{sec:numResults} probes.

The two parts of \cref{eq:dP2full} have a transparent origin at small modulation index. To second order in $\beta$, the drive moves a fraction $\beta^{2}/8$ of the mean power from the carrier into the two sidebands at $\bar\Delta\pm\Omega$, each carrying $\beta^{2}/16$. Keeping only the Bessel pairs $J_0J_{\pm1}$ in \cref{eq:Pi0}, and writing $T_k\equiv\chi_k\chi_{k-1}^{*}$,
\begin{equation}
  \Pi_0 \simeq \frac{\Lambda}{16}\,J_0J_1\bigl[3\,(T_0-T_1)+T_2-T_{-1}\bigr].
  \label{eq:Pi0-smallmu}
\end{equation}
To the same order the single-tone backaction harmonic is $\hat F_1^{(0)}(\bar\Delta)\simeq\Lambda J_0J_1(T_1-T_0)$. A tone shifted to $\bar\Delta\pm\Omega$ relabels every susceptibility, $\chi_k(\bar\Delta\pm\Omega)=\chi_{k\pm1}(\bar\Delta)$, and hence $T_k\to T_{k\pm1}$. Weakening the carrier by $\beta^{2}/8$ and adding tones at $\bar\Delta\pm\Omega$ of weight $\beta^{2}/16$ then changes the harmonic by
\begin{equation}
  \begin{aligned}
    &-\frac{\beta^{2}}{8}\hat F_1^{(0)}(\bar\Delta)
    +\frac{\beta^{2}}{16}\bigl[\hat F_1^{(0)}(\bar\Delta+\Omega)+\hat F_1^{(0)}(\bar\Delta-\Omega)\bigr]\\
    &\quad\simeq\frac{\beta^{2}\Lambda}{16}\,J_0J_1\bigl[-2(T_1-T_0)+(T_2-T_1)\\
    &\qquad\qquad+(T_0-T_{-1})\bigr]=\beta^{2}\,\Pi_0 ,
  \end{aligned}
\end{equation}
which is \cref{eq:Pi0-smallmu}. The phase-independent part of $\Delta P_2$ is therefore the dynamical backaction of the drive sidebands acting as additional pumps.
The sideband term also inherits the odd detuning dependence of single-tone backaction, vanishing at $\bar\Delta=0$, where tones placed symmetrically about the cavity resonance exert no backaction damping \cite{clerkBackactionEvasionSqueezing2008, kronwaldArbitrarilyLargeSteadystate2013, malzFloquetApproachBichromatically2016}. The phase-dependent part $-\tfrac12\Pi_0\,e^{+2i\phi}$ carries the same $\Pi_0$, so it is of the same order,
but varies with $2\phi$. It stems from the beat between the drive tones at $2\Omega$, which modulates the optical spring and damping parametrically \cite{rugarMechanicalParametricAmplification1991, mariGentlyModulatingOptomechanical2009, steinkeOptomechanicalBackactionevadingMeasurement2013, pontinSqueezingThermalMechanical2014, shomroniTwoToneOptomechanicalInstability2019}. Because every Bessel pair in \cref{eq:Pim} differs in index by three, $\Pi_-=\mathcal{O}(\mu^{3})$, so these two contributions make up all of $\Delta P_2$ for $\mu\lesssim1$. At larger $\mu$ the higher Bessel pairs mix the sidebands and the picture of independent pumps no longer holds.

\section{Injection locking and the locking range}
\label{supp:locking}

The main-text argument of \cref{eq:G-at-ABA} assumes the drive fixes the drive--velocity phase $\psi^\star$ on the energy-injecting branch, $\cos\psi^\star>0$. In the gray region of \cref{fig:enhancement}(d), where backaction alone does not reach threshold, there is no free-running oscillation for the drive to capture. Instead, the modulation itself generates the motion at $\omega_\mathrm{mod}$, so $\cos\psi^\star>0$ holds by construction. Only where backaction sustains a limit cycle of its own must the drive genuinely injection-lock it, which we test by comparing the optical-spring pull against the locking range. Injection locking of a radiation-pressure limit cycle to an intensity-modulated pump has been observed experimentally \cite{hossein-zadehObservationInjectionLocking2008, shlomiSynchronizationOptomechanicalCavity2015, arreguiInjectionLockingOptomechanical2021}. There, the locking region scales with the modulation force over the free-running amplitude \cite{shlomiSynchronizationOptomechanicalCavity2015}.

The free-running frequency is pulled away from $\Omega_m$ by the optical spring. For an oscillation with $n_\mathrm{cav}$ intracavity photons \cite{aspelmeyer2014review},

\begin{equation}
  \begin{split}
  \delta\Omega(\Delta) = g_0^2\, n_\mathrm{cav}
  \biggl[&
    \frac{\Delta+\Omega_m}{(\kappa/2)^2+(\Delta+\Omega_m)^2}\\
    &+\frac{\Delta-\Omega_m}{(\kappa/2)^2+(\Delta-\Omega_m)^2}
  \biggr],
  \end{split}
  \label{eq:optical-spring}
\end{equation}

which relies on the asymmetry between the two motional sidebands, like the backaction gain itself. Evaluated with the converged $n_\mathrm{cav}$ at every sweep point, the pull stays below $2.3\%$ of $\Omega_m$ at the most sideband-resolved, highest-power corner ($\kappa/\Omega_m=0.1$, $P_0=1\,\mathrm{mW}$), with a median of $9\times10^{-5}\,\Omega_m$.

The same first-order sideband beat that injects the power $\Delta P_1$ of \cref{eq:dP1-paper} also pulls the oscillator phase, through the other quadrature of the force harmonic. The component in phase with the displacement, $\mathrm{Re}\hat F_1=\beta\bigl[\mathrm{Re}(\Sigma_++\Sigma_-)\cos\phi-\mathrm{Im}(\Sigma_+-\Sigma_-)\sin\phi\bigr]$, does no work but shifts the oscillation frequency. It is a sinusoid in $\phi$ of amplitude $\beta K'/A\Omega$, with
\begin{equation}
  K'(A)=A\Omega\sqrt{\bigl[\mathrm{Re}(\Sigma_{+}+\Sigma_{-})\bigr]^{2}
  +\bigl[\mathrm{Im}(\Sigma_{+}-\Sigma_{-})\bigr]^{2}} .
  \label{eq:Kprime}
\end{equation}
Comparing with \cref{eq:K-paper} gives $K^2-K'^2=-4A^2\Omega^2\,\mathrm{Re}(\Sigma_+\Sigma_-)$, so $K'=K$ when $\Sigma_-$ is negligible. Because $\Sigma_-$ carries the Bessel pairs $J_nJ_{n\pm2}$, $|\Sigma_-/\Sigma_+|$ grows as $\mu^2$, and the two agree for $\mu\lesssim1$. Over the self-oscillating points of our sweep $K'/K$ has a median of $1.0$, but spans $0.4$--$6$ at $\mu\approx1$--$3$. The phase obeys the Adler equation, $\dot\vartheta=\delta\Omega-\Delta\Omega_\mathrm{lock}\sin\vartheta$, with half-width $\Delta\Omega_\mathrm{lock}=(\Omega_m/2Q_m)(A_\mathrm{inj}/A_\mathrm{free})$ \cite{adlerStudyLockingPhenomena1946}. A stationary locked phase exists whenever $|\delta\Omega|<\Delta\Omega_\mathrm{lock}$. The optomechanical version of this equation, with a locking strength scaling as the product of pump and drive amplitudes over the cycle amplitude, is derived in Ref.~\cite{amitaiSynchronizationOptomechanicalSystem2017}. A force of amplitude $F$ in phase with the displacement has $F=2\beta K'/A\Omega_m$ and resonant response $A_\mathrm{inj}=F/(m\Gamma_m\Omega_m)$. The amplitude ratio is therefore $A_\mathrm{inj}/A_\mathrm{BA}=\beta K'/P_\mathrm{diss}$, which reduces to the power ratio $\beta K/P_\mathrm{diss}$ plotted in \cref{fig:energy}(a) for small $\mu$. This gives
\begin{equation}
  \Delta\Omega_\mathrm{lock}
  = \frac{\Gamma_m}{2}\,\frac{\beta\,K'(A_\mathrm{BA})}{P_\mathrm{diss}(A_\mathrm{BA})} ,
  \label{eq:lockrange}
\end{equation}
with the locked phase $\sin\vartheta^\star=\delta\Omega/\Delta\Omega_\mathrm{lock}$, measured from the pulling quadrature. When $\Sigma_-$ is negligible, the pull and the injected power vary with the modulation phase $\phi$ as sinusoids a quarter period apart, so $\vartheta=\psi$: $\Delta P_1$ is maximal at the centre of the locking range and vanishes at its edges, where $\cos\psi^\star\to0$. For larger $\mu$, $\Sigma_-$ shifts these sinusoids relative to each other, so $\vartheta\neq\psi$ and $\Delta P_1$ no longer vanishes at the lock edge.

\begin{figure*}[t]
  \centering
  \includegraphics[
    width=0.8\figwidth
  ]{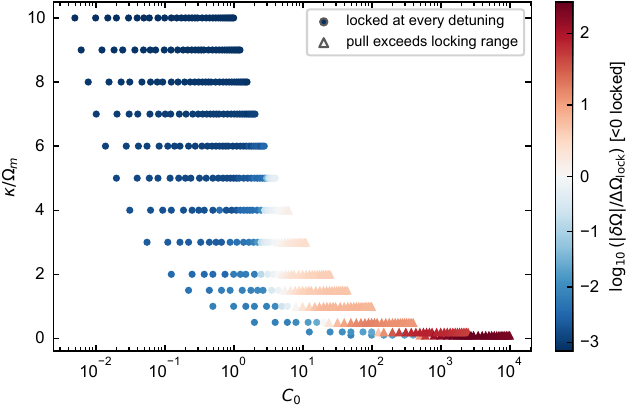}
  \caption{
    \textbf{Locking fails only where the modulation enhancement is
    smallest.}\\
    Worst-case ratio of the optical-spring pull $|\delta\Omega|$ (\cref{eq:optical-spring}) to the locking half-width $\Delta\Omega_\mathrm{lock}$ (\cref{eq:lockrange}) at $\beta=1$, maximized over detuning for each $(C_0,\kappa/\Omega_m)$. Circles mark combinations that lock at every detuning, so $\cos\psi^\star>0$; triangles mark combinations where the pull exceeds the locking range at some detuning. The color gives the ratio on a logarithmic scale (blue: locked, red: not locked). Only backaction-driven self-oscillating points are shown, since elsewhere no free-running frequency exists to capture. The failures lie at $C_0\gtrsim4$, $\kappa/\Omega_m\lesssim4$, the same corner where the enhancement $n_{*}/n_\mathrm{BA}$ in \cref{fig:enhancement}(d) is smallest.
    \label{fig:locking}
  }
\end{figure*}

In order to evaluate \cref{eq:lockrange}, we take $A_\mathrm{BA}$ from the converged $\beta=0$ amplitude. Evaluating \cref{eq:lockrange} at full depth for every self-oscillating point gives a median $\beta K'/P_\mathrm{diss}\approx2.2$ and half-width $1.1\times10^{-4}\,\Omega_m$, comparable to the median pull. Of all points, $59\%$ lock, and $64\%$ of $(C_0,\kappa/\Omega_m)$ combinations lock at every detuning. As shown in \cref{fig:locking}, the failures are confined to $C_0\gtrsim4$ and $\kappa/\Omega_m\lesssim4$, where the sideband asymmetry generating the spring is strongest. This is the same corner in which the enhancement is smallest (median $n_*/n_\mathrm{BA}\approx1.6$) and backaction already approaches the bandwidth-limited saturation. The assumption therefore weakens only where the enhancement is weakest. Where the enhancement is largest, and in the threshold-free region, $\cos\psi^\star>0$ holds. Evaluated at the locked phase itself, $\Delta P_1>0$ at $98\%$ of locked points; the exceptions lie near the lock edge at $\kappa/\Omega_m\le1$, $\mu\approx2$--$2.6$, within combinations that fail to lock at some detuning.

Two caveats bound this. First, $\beta K'/P_\mathrm{diss}\sim2$ sits beyond the edge of the small-signal regime in which the Adler form is derived. The strong-injection correction $(1-\rho^2)^{-1/2}$ \cite{kurokawaInjectionLockingMicrowave1973} broadens the range without altering this division. Second, $\delta\Omega$ is the linearized spring about $\bar x$. Neither affects the simulations of \cref{sec:numResults}, which integrates the full equations of motion with no locking assumption.

\bibliography{references}

\end{document}